\documentclass[prl,twocolumn,superscriptaddress,floatfix,amsmath,amssymb]{revtex4-2}
\usepackage[titletoc,toc,title]{appendix}
\usepackage{graphicx}
\usepackage{bm}
\usepackage{xcolor}
\usepackage{booktabs}
\usepackage{hyperref}
\usepackage{diagbox}
\usepackage{braket}
\usepackage{dcolumn}
\usepackage{gensymb}
\usepackage{array}
\renewcommand{\vec}[1]{\bm{#1}}
\usepackage{setspace}

\begin{document}

\title{
Geometric origins of topological insulation in twisted layered semiconductors
}

\author{Hao Tang}
\affiliation{Department of Physics, Peking University, Beijing, 100871, P.R.China}
\author{Stephen Carr}
\affiliation{Brown Theoretical Physics Center and Department of Physics, Brown University, Providence, Rhode Island 02912-1843, USA}
\author{Efthimios Kaxiras}
\affiliation{Department of Physics, Harvard University, Cambridge, Massachusetts 02138, USA}
\affiliation{John A. Paulson School of Engineering and Applied Sciences, Harvard University, Cambridge, Massachusetts 02138, USA}

\date{\today}

\begin{abstract}
Twisted bilayers of two-dimensional (2D) materials are proving a fertile ground for 
investigating strongly correlated electron phases.
This is because the moir\'e pattern introduced by the relative twist between layers 
introduces long-wavelength effective potentials which lead to electron localization.
Here, we develop a generalized continuum model for the electronic structure of moir\'e patterns,
based on first-principles calculations and tailored to capture the physics of twisted bilayer 2D semiconductors. 
We apply this model to a database of eighteen 2D crystals covering a range of atomic relaxation and electronic structure features. 
Many of these materials host topologically insulating (TI) moir\'e 
bands in a certain range of twist angles, which originate from the competition between triangular and hexagonal moir\'e 
patterns, tuned by the twist angle.
The topological phases occur in the same range as the maximally flat moir\'e bands.
\end{abstract}

\maketitle


Quantum materials, engineered by creative manipulation of 
the features of conventional crystals, offer 
new possibilities for breakthroughs in understanding electron correlations and superconductivity.
What is intriguing is that these phenomena emerge at a length scale 
much larger than the underlying crystal lattice constant (by a factor of 10 to 1000), 
through features due to strain patterns, controlled by geometric constraints.
A new platform for such studies are systems comprising few-layer
two-dimensional (2D) crystals, 
like twisted bilayers of graphene or transition metal dichalcogenides (TMDCs)~\cite{cao2018unconventional,wang2020correlated,yankowitz2019tuning,cao2018correlated,zondiner2020cascade}.
The slight lattice mismatch between two layers of a 2D 
material at a relative twist angle results in a moir\'e superlattice (MSL) 
and a long-wavelength periodic modulation of the effective electronic potential~\cite{santos2007tbg,park2008anisotropic}. 
In a narrow range of the twist angle, the moir\'e potentials act as confining wells for the electrons of the constituent monolayers, causing isolated flat bands and localized wave functions near the Fermi surface~\cite{bistritzer2011moire,morell2010flat}.

In the moir\'e flat bands,
the kinetic energy is heavily suppressed and electronic interactions play a dominant role, with the intensity of the interactions controlled by the twist angle; 
this effect has been dubbed ``twistronics''~\cite{carr2017twistronics}. 
Compared to twisted bilayer graphene (TBG), the twisted bilayer semiconductors can host flat bands in a large range of twist angles~\cite{xian2019multi,naik2018ultraflatbands} instead of at precisely a magic angle~\cite{cao2018unconventional}. This makes it possible to 
overcome some experimental challenges in twisted bilayers of semiconductors; thus, the twist angle becomes an additional degree 
of freedom for fine-tuning other physical effects in the strongly correlated regime~\cite{kerelsky2018magic}. 
Intriguingly, topological insulator (TI) moir\'e bands were predicted in a twisted bilayer 
of MoTe$_2$~\cite{wu2019topological}; this work 
introduced a possible candidate for observing concurrent correlated and TI phases in 
the same material.

It is difficult to model twistronic systems at small twist angles ($\theta \simeq 1\degree$) 
using first-principles calculations,  
because the number of atoms in the MSL scales as $\theta^{-2}$. To overcome this limitation, 
continuum models with a low-energy effective Hamiltonian based on first-principles 
(density-functional theory, DFT) 
calculations were developed for electronic structures of TBG; 
this approach can accurately describe flat bands and magic angles~\cite{carr2019exact,morell2010flat,jung2014ab,fang2016electronic}. 
Although continuum models have also been applied to the twisted bilayer semiconductors~\cite{wu2018hubbard,wu2019topological}, 
they have yet to include the effect of atomic relaxations which play an important role at small angles~\cite{nam2017tbh,carr2018relaxation,naik2018ultraflatbands,xian2019multi}.

In this letter, we present results from a DFT-based generalized continuum method designed specifically for 
twisted bilayer semiconductors. 
The computed electronic structures are consistent with full-DFT
results~\cite{naik2018ultraflatbands,xian2019multi,wang2020correlated} 
but only require a relatively inexpensive set of bilayer calculations. 
These calculations involve systems with only a handful of atoms per unit cell, in contrast to the many thousands of atoms
necessary in the calculation of a full MSL. We derive a database of relaxation and the  corresponding coefficients of tight-binding electronic structure hamiltonians for eighteen materials with various lattice symmetries and band edge momenta.
These coefficients do not capture the material's entire band structure, 
but rather focus on the details of the parabolic band edges.
Each layer contributes one band to the full twisted bilayer model, and 
these bands are coupled through a set of stacking-dependent electronic interactions.

\begin{figure}
\centering
\includegraphics[width=\linewidth]{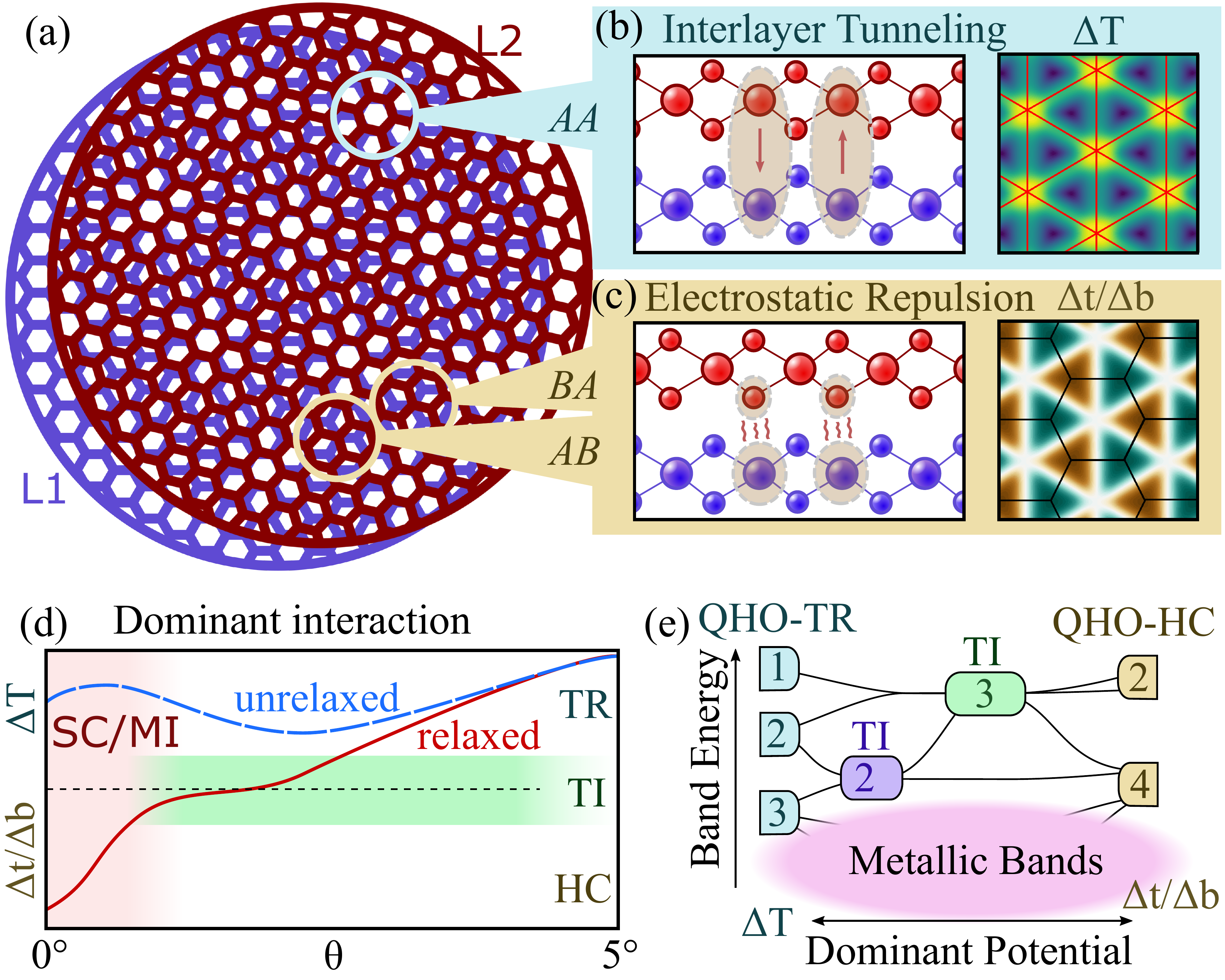} 
\caption{
(a) A moir\'e pattern of a twisted TMDC with areas of aligned ($AA$) stacking and eclipsed ($AB$ or $BA$) stacking highlighted.
(b) At $AA$ stacking, the local electronic Hamiltonian is described by interlayer tunneling between the layers ($\Delta T$), forming a triangular (TR) lattice.
(c) At $AB$ or $BA$ stacking, the electronic Hamiltonian is  described by interlayer electrostatic potentials ($\Delta t, \Delta b$), forming a honeycomb (HC) lattice.
(d) Dependence of the dominant moir\'e interaction on twist angle,  with atomic relaxation of the superlattice (red line) or without it (blue line): evolution of the lattice character from TR to HC with {\em decreasing} $\theta$ induces TI states and flat bands at small angles, leading to Mott insulator (MI) and superconducting (SC) behavior.
(e) The TI phase is caused by band hybridization as the electrons transition between quantum harmonic oscillator (QHO) states on a TR to a HC lattice.}
\label{fig:intro_fig}
\end{figure}

In Fig.~\ref{fig:intro_fig} we provide an overview of the different phases and their geometric 
origins, as the twist angle is changed in a moir\'{e} bilayer (Fig.~\ref{fig:intro_fig}a).
The full interaction between the bands can be decomposed into two complementary parts.
The first describes the tendency for electrons in one layer to tunnel to the other, 
as shown in Fig.~\ref{fig:intro_fig}b -- labeled $\Delta T$ for ``Tunneling''. 
The tunneling coefficients are strongest at $AA$ (aligned) stacking regions 
and form a triangular (TR) lattice across the MSL.
The second contribution captures the stacking dependence of the monolayer bands' 
on-site energies, which depends on the electrostatic potential from the opposite 
layer, shown in Fig.~\ref{fig:intro_fig}c) -- labeled $\Delta t$ and $\Delta b$ for ``top/bottom'' layer.
The electrostatic potentials have maxima at the $AB$ and $BA$ 
stacking regions, forming a honeycomb (HC) lattice. 

\renewcommand{\baselinestretch}{1.0}

In the majority of materials studied here, the in-plane atomic relaxation of the twisted bilayers 
enhances the effects of the electrostatic potential fluctuations 
and reduces that of the interlayer tunneling, especially at small twist angles.
Electron localization transitions, from the tunneling-dominated TR lattice to the electrostatic-dominated HC lattice, 
occur at low twist angle in various materials, as illustrated in Fig.~\ref{fig:intro_fig}d. 
In the low-angle region, ultra-flat moir\'e bands are predicted, likely to host superconducting (SC) and Mott insulating (MI) phases. 
The topologically insulating (TI) moir\'e bands appear for intermediate values of the twist angle, 
during the transition between the two types of lattice geometries for the interlayer 
electronic interactions (Fig.~\ref{fig:intro_fig}e). 
The electronic potentials naturally lead to quantum harmonic oscillator (QHO) states 
at small twist angles~\cite{carr2020duality, angeli2020gamma, naik2020dots}, 
whose presence underlies the appearance of the moir\'e flat bands 
and whose competition explains the large number of TI phases possible in twisted semiconductors.

We briefly introduce the methodology here. 
To capture the stacking-dependent electronic and atomic details,
we perform DFT calculations on aligned bilayers over a grid sampling all  possible interlayer displacements.
Using the effective mass approximation, we treat the dispersion around the monolayer band extrema as a kinetic energy term in a continuum Hamiltonian.
Including the three stacking-dependent potentials described in the introduction, we obtain the effective Hamiltonian:
\begin{equation}
H=\left( \begin{matrix} 
-\frac{\hbar^2 (\nabla -ik_0)^2}{2m^*}+\Delta_t(\vec{r}) & \Delta_T^*(\vec{r}) \\ \Delta_T(\vec{r}) &-\frac{\hbar^2(\nabla -ik_0)^2}{2m^*}+\Delta_b(\vec{r}) 
\end{matrix}\right)
\label{eq:cont_ham}
\end{equation}
where $m^*$ is the effective mass, $\Delta_{t,b}$ are the electrostatic potentials 
for the top and bottom layer, and $\Delta_T$ is the interlayer tunneling strength. 
At small twist angles
we also include the all-important relaxation effects by minimizing the total mechanical energy of the moir\'e patterns \cite{carr2018relaxation}.
The local electronic structures can then be derived from the DFT through the expansion:
\begin{equation}
E^{(\pm)}(\vec{r},\vec{k})=E^{(\pm)}(\vec{\theta}\times \vec{r}+2\vec{u})+\frac{\hbar^2 (\vec{k}-\vec{k}_0)^2}{2m^*}
\end{equation}
To determine the $\Delta_{t/b},\Delta_T(\vec{r})$ from the local electronic structure, the Bloch wave functions 
at band extrema are extracted from DFT calculations to assess the layer polarization of each band. The moir\'e bands can then be calculated by diagonalizing the matrix form of the Hamiltonian for a truncated basis set in $k$-space (see SM for details).

The aligned ($AA$) stacking configuration has higher energy 
than the partially eclipsed ($AB$/$BA$) configuration for TMDCs and hBN 
homo-bilayers ($\theta\simeq 0\degree$).
Consequently, relaxation tends to reduce the in-plane area of $AA$ stacking region 
and to increase that of the $AB$/$BA$ stacking region, thus minimizing the total 
energy~\cite{jung2014ab,carr2018relaxation}. 
Upon relaxation, the large values of $\Delta_{t,b}$, at $AB$ and $BA$ stacking, expand to cover a larger area, while the peak regions of $\Delta_T$, at $AA$ stacking, shrink, 
with relaxed structures showing stronger electrostatic potential effects and weaker tunneling effects, 
which causes a clear 
angle-dependent transition of the moir\'e electronic structure.

\begin{figure}[hbtp]
\centering
\includegraphics[width=\linewidth]{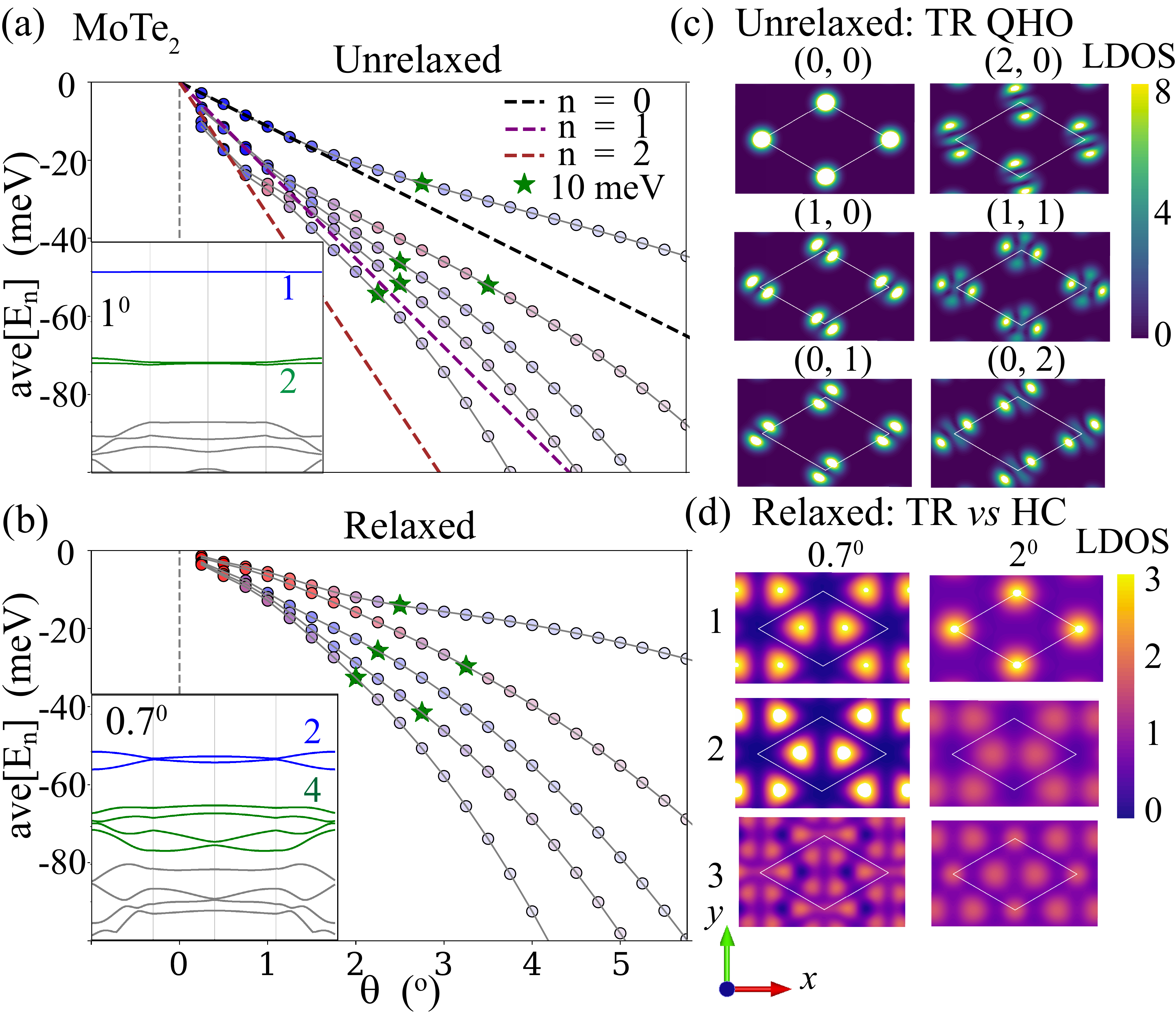}
\caption{Angle-dependent band structures and real-space localization. 
(a, b) The band average energy (ave[E$_\textrm{n}$]), bandwidth (color saturation, deeper color 
for flatter bands), and localization (red for $AB$/$BA$ stacking and blue for $AA$ stacking) 
of the three top moir\'e valence bands as a function of the twist angle $\theta$ 
for twisted bilayers of MoTe$_2$. A fitted QHO model is given by the dashed lines, 
and the first twist angles with bandwidths less than 10 meV are denoted with green stars. 
The $1\degree$ band structures are shown in the insets, with the first two QHO levels highlighted in color. 
(c) LDOS of the top six moir\'e bands of unrelaxed bilayer MoTe$_2$ twisted at 0.5$\degree$. 
The quantum numbers in the QHO model are denoted as n = ($n_x,n_y$). (d) LDOS of the moir\'e bands 
for a 0.7$\degree$ and a 2$\degree$ twisted bilayer of MoTe$_2$. 
The average LDOS in the MSL are normalized to one in both (c) and (d).}
\label{fig:localization}
\end{figure}

In Fig.~\ref{fig:localization}a,b we present the angle-dependent bands and real-space localization 
of twisted bilayer MoTe$_2$ as a representative case. 
The top-most valence bands for both the relaxed and unrelaxed moir\'e 
systems have a bandwidth less than 10 meV when the twist angle is below $2.5\degree$.
Using the Coulomb repulsion energy ($U$) in TBG and twisted hBN as a 
guide~\cite{cao2018correlated,xian2019multi}, this small kinetic energy implies that strongly 
correlated states could exist for any twist angle below a certain critical value in some TMDC
materials~\cite{wang2020correlated,zhang2020tmdcs}.
The relaxation effects drive the small-angle ($<2\degree$)
valence moir\'e bands from TR-type to HC-type (Fig.~\ref{fig:localization}).
In the unrelaxed case, the band structure at $1\degree$ shows a single uppermost flat band 
and a pair of flat bands under it, consistent with a $AA$ 
stacking QHO model~\cite{carr2020duality,angeli2020gamma}. 
The energy levels of the 2D TR QHO model are given by $E_n =- \hbar\omega(n_1+n_2+1)$, 
$\omega=\omega_{\theta}\theta$, and the first three levels $-\hbar\omega$,$-2\hbar\omega$, 
and $-3\hbar\omega$ are accurately represented by the computed band averages for 
$\theta<1\degree$ \cite{carr2020duality}.
In contrast, a pair of top moir\'e bands and the four lower bands at $0.7\degree$ correspond to the first ($n=0$) and second ($n=1$) QHO states, respectively of the HC lattice \cite{angeli2020gamma}. 

The real space localization of the tunneling-dominated (TR) QHO state is evident 
in the local density of states (LDOS) of the moir\'e bands for the unrelaxed condition, 
as shown in Fig.~\ref{fig:localization}c. 
The electronic states localize around the $AA$ stacking center and show 
\textit{s}, \textit{p}, and \textit{d}-orbital distribution for the first three energy levels, respectively.
The six moir\'e bands shown correspond to the QHO ground state, single excitation states 
(2-fold degenerate), and double excitation states (3-fold degenerate). 

\begin{figure*}[hbtp]
\centering
\includegraphics[width=\linewidth]{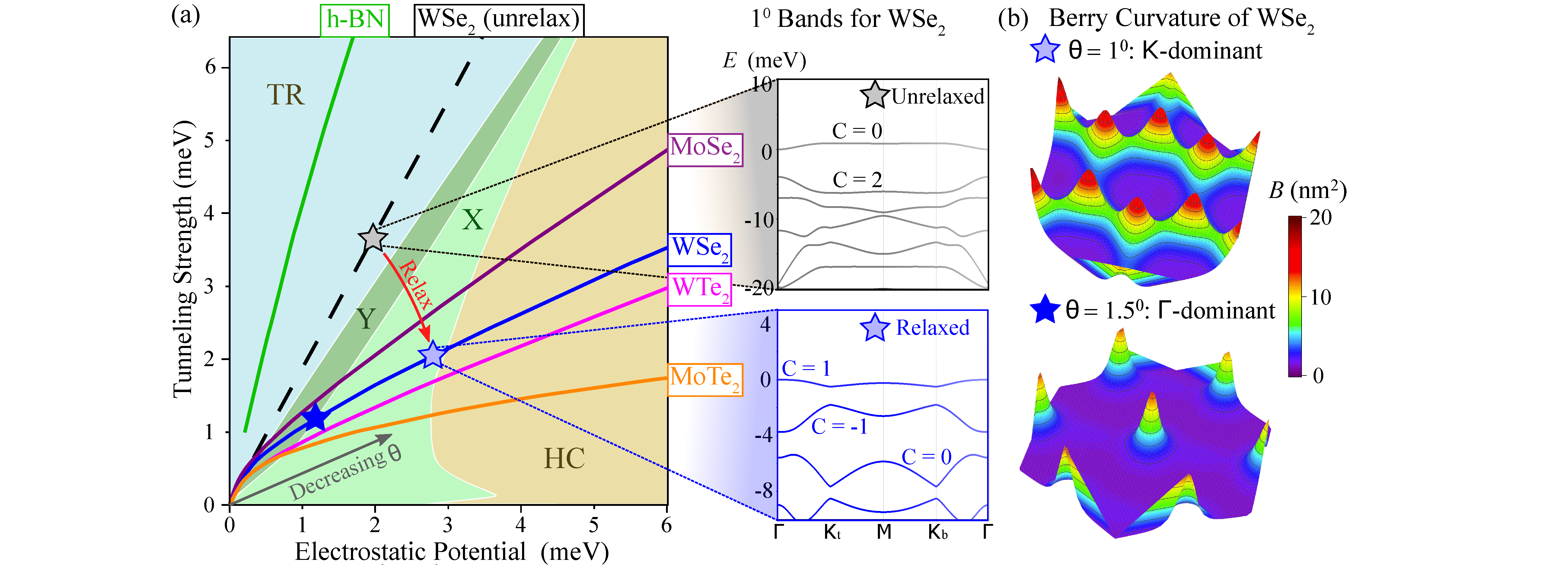}
\caption{Topological phase diagram for moir\'e bands. 
(a)
Phase of the moir\'e bands as a function of the intensity of potential and tunneling fluctuations 
(1st Fourier components). The light (dark) green areas represent X (Y) type 
TI phase, while the orange and light blue areas represent topologically trivial HC and TR QHO states, 
respectively. The twist-angle dependence of the relaxed band structures is shown by the solid lines for 
selected materials. The band structures and Chern numbers for unrelaxed and relaxed moir\'e bands in 
$1\degree $ twisted bilayer WSe$_2$ are compared. (b) Berry curvature of the uppermost band for X-type 
($1\degree $) and Y-type ($1.5\degree$) twisted bilayer WSe$_2$ in the MBZ.}
\label{fig:TI}
\end{figure*}

After relaxation, the competition between the two types of lattices is greatly altered, 
as shown in Fig.~\ref{fig:localization}d.
The two top bands in the low-angle case ($\theta = 0.7\degree$) localize in the $AB$/$BA$ HC potential wells to from 2-fold degenerate ground states of the HC QHO model, 
while the third band is one of the 4-fold degenerate first excited states. 
In contrast, the localization of the uppermost band gradually changes to the $AA$ stacking in the high-angle case ($\theta = 2\degree$) with the second band localized in the $AB$/$BA$ region. 
This indicates that the energy level of the HC QHO model and the triangular QHO model 
cross as the angle changes, leading to a reordering of the bands’ localization. 
Therefore, the transition between the HC and TR electronic states can be controlled 
by varying the twist angle in the presence of atomic relaxations. 
The intermediate twist angles correspond to competing real-space distribution 
and band reordering, making these materials excellent candidates for hosting 
non-trivial topological properties.

 To study the topological properties associated with the TR-HC transition, 
 we calculate the Chern numbers of the moir\'e bands for a generic twisted bilayer 
 of a hexagonal semiconductor, and use these results to generate a TI phase diagram. 
 The condition for the existence of topological bands is represented by the electrostatic potential 
 and tunneling coefficients in Fig.~\ref{fig:TI}a for a generic system, which are approximated by their first order Fourier coefficients labeled $V$ and $w$, respectively. 
 In this simplified model, the topological properties
 can be completely described by just two parameters, 
 $\alpha = Vm^*a^2/\theta^2$ and $\beta = wm^*a^2/\theta^2$, with the Hamiltonian

\begin{equation}
H \approx \hat{K}+\hat{V}+\hat{W} = \frac{\theta^2}{m^*a^2} \left( \hat{K_0}+\alpha \hat{V_0}+\beta \hat{W_0} \right)
\label{eq:V_w_ham}
\end{equation} 
where $a$ is the lattice parameter, $\hat{K}$, $\hat{V}$, $\hat{W}$ are the kinetic, potential, 
and tunneling energy, respectively, and $\hat{K_0}$, $\hat{V_0}$, $\hat{W_0}$ are 
the material-independent 
quantities.
The topologically trivial TR and HC phases appear in the $w$-dominated and $V$-dominated regions 
of the phase space of Fig.~\ref{fig:TI}a, respectively. Interesting band structures occur in the intermediate
region, which includes two types of topological non-trivial valence moir\'e bands. The Chern numbers of the 
three top-most moir\'e bands are (1, -1, 0) and (1, 1, -2) for X-type and Y-type topological insulators, 
respectively. Transitions between topological and trivial insulators for the uppermost band will occur on the 
boundary between the Y-type topological phase and TR phase where the top two bands overlap and then 
reopen, hosting protected edge modes at the transition. Similar topological transitions between the second 
and the third bands appears in the boundary between the X-type and Y-type topological phases. The transition 
on the boundary of the X-type topological phase and the HC phase is a TI-to-semimetal transition.

In Fig.~\ref{fig:TI}a we show the curves in the phase diagram for 4 TMDC's, indicating transitions 
between phases with increasing twist angle (without relaxation these curves would be straight lines like the one shown for WSe$_2$).  For comparison, we include the curve for hBN, which remains in one phase (TR) as it shows very weak relaxation. 
An explicit example of the topological transitions is also shown Fig.~\ref{fig:TI}a for WSe$_2$, including the unrelaxed case. 

The band structures and Chern numbers of the bands confirm that the topological transition is 
primarily driven by atomic relaxation. 
In the unrelaxed 1$\degree$ twisted WSe$_2$, although the uppermost valence band is topologically trivial, the second valence band has a nonzero Chern number. 
The topological features of these lower valence bands are not reflected by the phase diagram, but we list them in our database for each material.
In realistic experimental conditions, atomic-scale defects, bending, and local strain introduced during 
fabrication cause twisted bilayers to include different twist angles in different areas within 
a single sample \cite{uri2020squid}. Therefore, different domains of twist angle occur and introduce both 
topological and trivial bands. Topological edge states can appear along domain boundaries in the twist angle, 
according to the critical values in Tab.~\ref{tab:crit_values}.
These edge states would not be defined by a sharp change from one crystal to another, or from material to vacuum, but 
rather by slow variation in the twist angle.
For this reason, these ``internal'' protected edge states would be excellent candidates for observing spin or 
valley-polarized states, as the only disorder comes from twist-angle variations, 
which are unlikely to induce a spin or valley swapping.

\begin{table}[h]
\centering
\caption{
$\theta_{\rm HC/X}, \theta_{\rm X/Y},$ and $\theta_{\rm Y/TR}$ are the critical twist angles (in degrees) for transitions of the top bands between 
the indicated phases.
$E_{\rm gap}$ is the maximum band gap 
between the top and lower bands (in meV), and $\omega_{\rm top}$ is the minimum band width of the top bands in the 
topologically non-trivial regime (in meV); 
$\theta_{\rm gap}$ and $\theta_{\rm top}$ are the angles where these extrema occur.}
\begin{tabular}{|c|c|c|c|c|c|}
\hline 
\rule[-1ex]{0pt}{1ex} Material & $\theta_{\rm HC/X}$ & $\theta_{\rm X/Y}$ & $\theta_{\rm Y/TR}$ & $E_{\rm gap}(\theta_{\rm gap})$ & $\omega_{\rm top}(\theta_{\rm top})$\\ 
\hline 
\rule[-1ex]{0pt}{1ex} MoSe$_2$ & 0.9 & 1.4 & 2.4 & 2.98 ($1.67 \degree$) & 0.18 ($1.21\degree$) \\ 
\rule[-1ex]{0pt}{1ex} MoTe$_2$ & 0.7 & 1.1 & 1.5 & 1.43 ($1.20\degree$) & 0.30 ($1.10 \degree$)\\ 
\rule[-1ex]{0pt}{1ex} WSe$_2$ & 0.8 & 1.2 & 2.1 & 3.24 ($1.40\degree$) & 0.19 ($1.15\degree$) \\ 
\rule[-1ex]{0pt}{1ex} WTe$_2$ & 0.7 & 1.4 & 2.4 & 3.70 ($1.68 \degree$)& 0.18 ($1.31 \degree$)\\  
\hline 
\end{tabular}
\label{tab:crit_values}
\end{table}

In Fig.~\ref{fig:TI}b we show the Berry curvature 
of the uppermost valence band for two values of the twist angle. In the HC phase in the small-angle region, the first transition 
appears at $\theta_{\rm HC/X} = 1\degree$ where the two top bands separate at the MBZ's $K$ points, indicated by the 
Berry curvature's concentration there.
With increasing twist angle, the Berry curvature of the top band gradually transfers to the $\Gamma$ point accompanied by an X/Y transition at
$\theta_{\rm X/Y} = 1.5\degree$. During this process, the topologically non-trivial top-most band takes on its minimal 
bandwith , $\omega_{\rm top} \simeq 0.19$ meV, and its maximal topological band gap 
$E_{\rm gap} \simeq 3.24$ meV. At larger twist angle, the Berry curvature concentrates at the 
$\Gamma$ point during the transition to the TR phase, where the two top-most bands merge. The numerical 
results for the critical angles, maximal gap, and minimal band width for the TMDCs with topological valence 
bands are presented in Table 1. 
We have also verified that the bands and 
real space localization of MoS$_2$~\cite{naik2018ultraflatbands}, WSe$_2$~\cite{wang2020correlated}, 
and hBN~\cite{xian2019multi} from our model are consistent with previous full DFT calculations.
Twisted bilayers of WTe$_2$ show the highest gap and one of the lowest bandwidths, 
making this material the best candidate for experimental study. 

Including spin degrees of freedom,
the spin-dependent moir\'e Hamiltonian decomposes into two copies \cite{wu2019topological}.
At the $K$ valleys of the aligned bilayer, the two copies will be split by
the spin-orbit coupling term, 
$\Delta_\textrm{SOC}$, which for most
TMDCs is on the scale of $100$ meV, and the opposite spin-ordering will occur at the $K'$ valley.
The $\Gamma$ and $M$ points tend to have very weak spin-splitting, leading to two copies of the spin-independent 
Hamiltonian, as is the case in continuum 
treatments of twisted bilayer graphene~\cite{bistritzer2011moire}).
However, as the two nonequivalent $K$ valleys are related by time reversal symmetry, the spin-up bands 
at $K$ valley and spin-down bands at $K'$ valley have opposite Berry curvature and Chern numbers. 
Topologically non-trivial uppermost bands with opposite spin and Chern number give rise to the helical edge 
states protected by the TR symmetries at the boundaries, and could lead to observable quantum spin Hall 
(QSH) effects~\cite{RevModPhys.83.1057}.

The origin of the TI phases has been ascribed~\cite{wu2019topological} 
to a skyrmionic texture in the moir\'e $\Delta$ potentials, which is an alternate interpretation of the TR/HC competition we presented here but captures the same key 
features. As the flatness of the bands causes stronger correlation effects, the topological 
phases in twisted bilayer semiconductors could exhibit the combination of TI and superconductivity,
which has been the subject of an intense decades-long search for fractional statistics 
and Majorana fermions~\cite{RevModPhys.83.1057}.

The generality of the relaxation-induced TI phases is partially explained by the connection between the 
stacking-dependence of the electronic structure and that of the ground state energy: the lowest stacking 
configuration of these materials seem to correspond to maxima 
of the electrostatic interaction and minima of the tunneling strength. 
We note that this trend might not be true across all 2D materials, 
and finding exceptions to the rule could be a valuable endeavor.

\begin{acknowledgments}
We thank Ziyan Zhu and Daniel Larson for useful discussions. This work was supported by the National Science Foundation under grant No. OIA-1921199. The calculations in this work were performed in part on the FAS Research Computing cluster of Harvard University.
\end{acknowledgments}

\bibliography{refs.bib}
\clearpage

\end{document}


\title{
Geometric Origins of topological insulation in twisted semiconductors \\(supplementary materials)
}

\author{Hao Tang}
\affiliation{Department of Physics, Peking University, Beijing, 100871, PRC}
\author{Stephen Carr}
\affiliation{Brown Theoretical Physics Center and Department of Physics, Brown University, Providence, Rhode Island 02912-1843, USA}
\author{Efthimios Kaxiras}
\affiliation{Department of Physics, Harvard University, Cambridge, Massachusetts 02138, USA}
\affiliation{John A. Paulson School of Engineering and Applied Sciences, Harvard University, Cambridge, Massachusetts 02138, USA}
\date{\today}

\maketitle
\section{Methodology}\label{sec:method}
The DFT calculations of commensurate bilayers use the projector-augmented wave (PAW) basis sets with cut-off energy of 400 eV implemented by the Vienna ab initio simulation package (VASP)~\cite{PhysRevB.54.11169,PhysRevB.59.1758}. To overcome the band gap problem of the generalized gradient approximation (GGA), we perform a meta-GGA calculation with Strongly constrained and appropriately normed (SCAN) semilocal density functional~\cite{PhysRevLett.115.036402,sun2016accurate}, with the vdW-DF functional rVV10 to more accurately capture the van der Waals interaction~\cite{PhysRevX.6.041005}. 
A vacuum layer of 15 \AA{} in the c ($z$) direction is set and the dipole corrections are switched on to cancel any residual inter-slab interactions. The k-point mesh is sampled by the Monkhorst package with separation of 0.2 \AA{}~\cite{PhysRevB.54.11169}. All the electronic iterations are converged to $10^{-5}$ eV and the atomic relaxations are converged to $10^{-4}$ eV. 

The mechanical properties (generalized stacking fault energy (GSFE))~\cite{carr2018relaxation} and electronic structure properties are calculated for displacement $\vec{d}=u_1\vec{a_1}=u_2\vec{a_2}$ on a $9\times 9$ grid in the unit cell for each bilayer material, where $a_{1,2}$ is the unit cell lattice vectors. For each displacement, the in-plane position of the atoms is fixed while the layer distance is relaxed. The typical variation ranges of $z$ coordinates of TMDCs for different stacking are about 0.4 \AA{}. The strain modulus, GSFE, band edge momenta, energy level, effective mass, the corresponding Kohn-Sham wave functions, and their projection weight to the top and bottom layers, are extracted for each atomic displacement independently, and then applied in the continuum model.

\begin{figure}[hbtp]
\centering
\includegraphics[width=\linewidth]{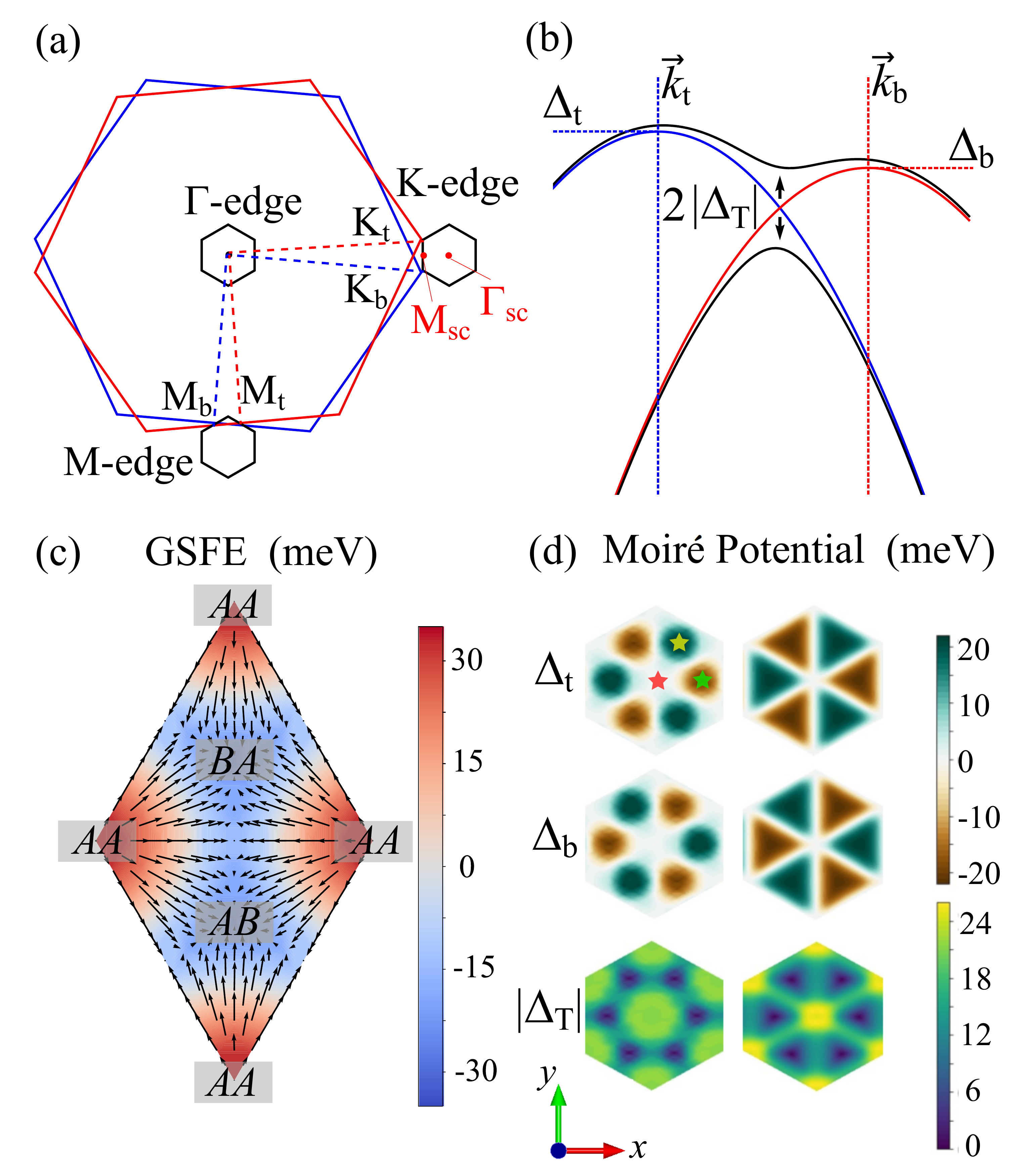}
\caption{
Electronic model for the twisted bilayers.
(a) Brillouin zones for a twisted bilayer 
of hexagonal lattices. The black hexagons show the MBZ expanded around selected high symmetry points. 
(b) Formation of moir\'e flat bands (black) from top (red) and bottom (blue) single-layer bands near the band 
(c) GSFE landscape (color) and atomic displacements (arrows) for a 2$\degree$ twisted bilayer of WSe$_2$. 
(d) $\Delta_{t,b}$ and $|\Delta_T|$ values from DFT without (left) 
and with (right) atomic relaxation; stars 
represent $AA$ (red), $AB$ (green), and $BA$ (yellow) stacking.}
\label{fig:dft_methods}
\end{figure}

The continuum model for the moir\'e superlattice (MSL) of twisted bilayers is described as follow. The translation symmetry of the moir\'e superlattice defines the moir\'e Brillouin zone (MBZ), 
as illustrated in Fig.~\ref{fig:dft_methods}a.
The moir\'e bands closest to the band gap can be derived from the band extrema 
in the original Brillouin zone, which are typically located at high symmetry $k$-points. 
In the local electronic structures, the bands 
of individual layers are shifted by the electrostatic potential $\Delta_{t,b}$ and split by the magnitude 
of the interlayer tunneling $2|\Delta_T|$, as shown in Fig.~\ref{fig:dft_methods}b. The electronic Hamiltonian in the MSL is then constructed by these potentials as Eq.~(1) in the main body.

At small twist angles
we also include the all-important relaxation effects by minimizing the total mechanical energy of the moir\'e patterns \cite{carr2018relaxation}.
This energy is a summation of the (GSFE) 
and intra-layer strain energy:
\begin{subequations}
\begin{eqnarray}
E = \sum_{j=t,b}\int \frac{1}{2}\epsilon_j C_j \epsilon_j d^2\vec{r}
+\int V_{\rm{GSFE}}(\vec{d}(\vec{r}))d^2\vec{r}
\\
\vec{d}(\vec{r}) = 
\vec{d_0}(\vec{r})+\vec{u_t}(\vec{r})-\vec{u_b}(\vec{r})
\end{eqnarray}
\end{subequations}
where $\vec{u}_{t/b}(\vec{r})$ is the in-plane displacement of the top/bottom layer, 
and $\epsilon$, $C$, $V_{\rm{GSFE}}$, $d_0$ are the strain tensor, 
strain modulus matrix, GSFE density, and displacement under rigid twist, respectively.
The effect of atomic relaxations can be easily incorporated into the electronic Hamiltonian. In the unrelaxed case, the DFT-extracted values of the stacking-dependent potentials are 
unwrapped onto the moir\'e pattern with a straightforward linear mapping.
To include relaxations we add one step after the linear mapping 
which updates the spatial dependence of the potentials to accurately reflect
the relaxed configurations of the bilayer.
As the stacking displacement is a function of the real space coordinate,
$\vec{d}(\vec{r})=\vec{\theta}\times \vec{r}+2\vec{u}$ 
($\vec{u}=\vec{u}_t=-\vec{u}_b$, considering the layer-exchange symmetry for the in-plane relaxations), we derive the local band energy at arbitrary position $\vec{r}$ expanded in the effective
mass approximation:
\begin{equation}
E^{(\pm)}(\vec{r},\vec{k})=E^{(\pm)}(\vec{\theta}\times \vec{r}+2\vec{u})+\frac{\hbar^2 (\vec{k}-\vec{k}_0)^2}{2m^*}
\end{equation}
where the band extrema energy $E^{\pm}$ and effective mass $m^*$ are determined by the DFT.

To determine the $\Delta_{t/b},\Delta_T(\vec{r})$ from the local electronic structure, the Bloch wave functions 
at band extrema are extracted from DFT calculations to assess the layer polarization of each band. 
At the band extrema of bilayer structures, any non-degenerate monolayer band edge splits 
into the bonding ($+$) and antibonding ($-$) combinations, with energies $E^{(\pm)}$ and 
wavefunctions $\psi^{(\pm)}_k(r)$, containing contributions from the top ($\phi^t_k(r)$) and bottom ($\phi^b_k(r)$) layers, with coefficients $c^{\pm}_{t/b}$ obtained from DFT calculations:

\begin{figure*}[hbtp]
\centering
\includegraphics[scale=0.22]{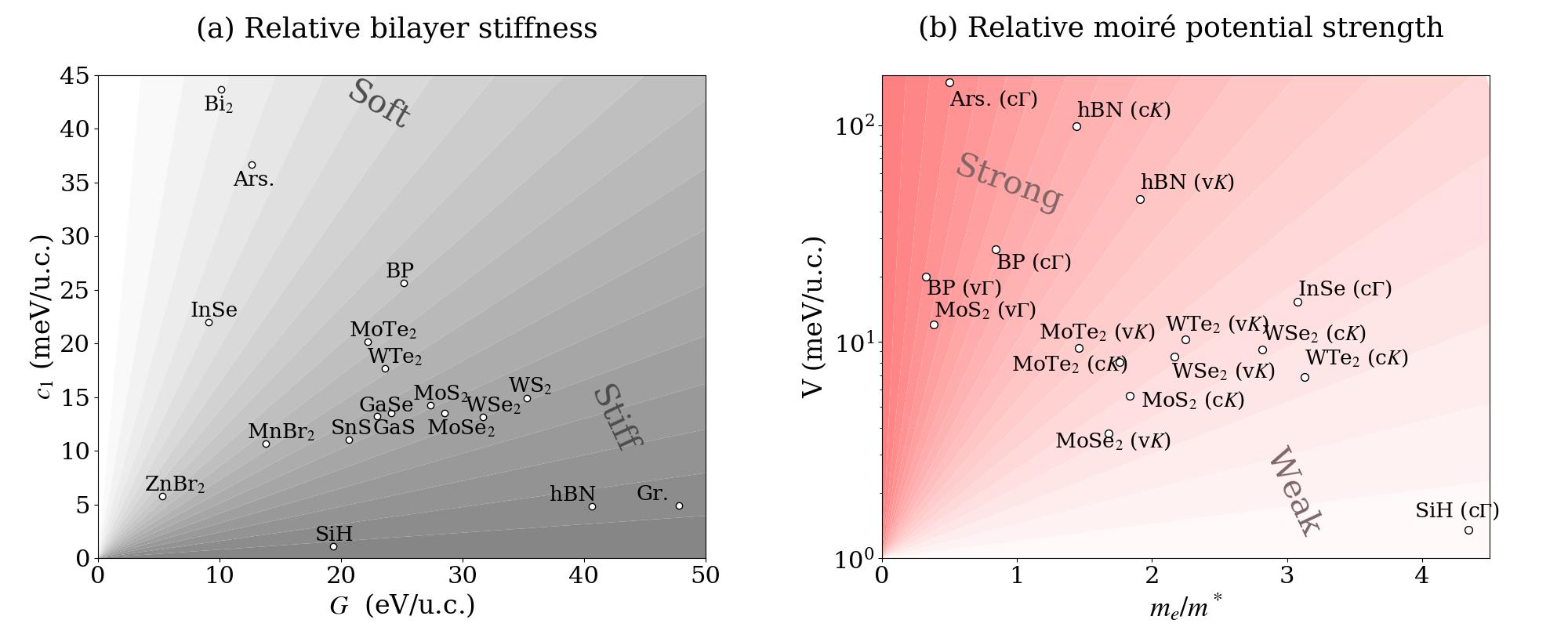}
\caption{Comparison of relaxation and electronic coefficients. (a) First order Fourier components of the GSFE energy $c_1$ $vs$ anisotropic strain coefficient G. The relative intensity of intralayer and interlayer energy coefficients from stiff to soft is remarked by the color from dark to bright. (b) Inverse effective mass vs the range of potential energy fluctuation in the MBZ. The relative importance of energy terms is remarked by white and red color for kinetic energy dominant and potential energy dominant condition, respectively. }
\label{fig:matlist}
\end{figure*}
\begin{table*}
\caption{Strain and GSFE coefficients and critical twist angles for atomic relaxation. The coefficients K and G are strain modulus for isotropic and anisotropic distortion with the unit of eV per unit cell, respectively. $c_1$ is 2 times of the 1$^{st}$ order Fourier component of GSFE as defined in ref.\cite{carr2018relaxation}. The critical twist angle $\theta_c$ for evident relaxation effect is evaluated as equ.~\ref{equ:ang_rel}, with regarding the critical twist when the separate domain walls begin to appear.}
\begin{tabular}{|c|c|c|c|c|} 
\hline
\ \ \ Materials \ \ \ & \ $ G $ (eV/u.c.)\ &\ $ K $ (eV/u.c.)\ & \ $ c_1 $ (meV/u.c.)\ & \ \ \ $\theta_c (^0)$ \ \ \  \\ 
\hline 
MoS$_2$ & 27.36 & 45.45 & 14.25 & 1.7 \\ 
MoSe$_2$ & 28.52 & 43.17 & 13.54 & 1.7 \\ 
MoTe$_2$ & 22.16 & 35.63 & 20.18 & 2.2 \\ 
WS$_2$ & 35.27 & 52.33 & 14.93 & 1.6 \\ 
WSe$_2$ & 31.67 & 46.22 & 13.11 & 1.6 \\ 
WTe$_2$ & 23.61 & 33.82 & 17.67 & 2.1 \\ 
hBN & 40.66 & 59.65 & 4.82 & 0.8 \\ 
SiH & 19.33 & 31.47 & 1.13 & 0.6 \\ 
Ars. & 12.62 & 36.04 & 36.71 & 3.0 \\ 
BP & 25.14 & 43.23 & 25.67 & 2.3 \\ 
Bi$_2$ & 10.15 & 16.81 & 43.71 & 4.8 \\ 
InSe & 9.1 & 25.47 & 22.01 & 2.8 \\ 
Gr. & 47.78 & 67.87 & 4.9 & 0.8 \\ 
GaS & 24.11 & 38.9 & 13.53 & 1.7 \\ 
SnS & 20.62 & 34.1 & 11.04 & 1.7 \\ 
MnBr$_2$ & 13.79 & 20.41 & 10.68 & 2.1 \\ 
ZnBr$_2$ & 5.3 & 20.93 & 5.74 & 1.6 \\ 
GaSe & 22.98 & 36.67 & 13.2 & 1.8 \\ 
\hline 
\end{tabular}
\label{tab:relax}
\end{table*}
\begin{gather}
\psi_{k}^+(r)=c_b^+ \phi_{k}^b(r)+c_t^+ \phi_{k}^t(r) \\
\psi_{k}^-(r)=c_b^- \phi_{k}^b(r)+c_t^- \phi_{k}^t(r)
\end{gather}
The coefficients $c_{t,b}^{\pm}$ satisfy the eigenvalue equation of local electronic structures:
\begin{equation}
\left[
\begin{matrix}
\Delta_b &\Delta_T\\
\Delta_T^* &\Delta_t
\end{matrix}\right]\left[
\begin{matrix}
c_b^{\pm} \\
c_t^{\pm}
\end{matrix}\right]=E^{\pm}\left[
\begin{matrix}
c_b^{\pm} \\
c_t^{\pm}
\end{matrix}\right].
\end{equation}
Consequently, $\Delta_t, \Delta_b$ and $\Delta_T$ can be solved for at each displacement as follows:
\begin{equation}
\Delta_t=\frac{|c_t^+|^2E^++|c_b^+|^2E^-}{|c_t^+|^2+|c_b^+|^2},    \Delta_b=\frac{|c_b^+|^2E^++|c_t^+|^2E^-}{|c_t^+|^2+|c_b^+|^2}
\end{equation}
\begin{equation}
\Delta_T=\frac{(c_t^+)^*c_b^+}{|c_b^+|^2+|c_t^+|^2}(E^+-E^-)
\end{equation}

The $c$ coefficients can be derived from DFT calculations by two methods.
In the first method, one projects the Bloch wave functions onto atomic orbitals in the top/bottom layer, so that the value of $|c_{t/b}^{\pm}|$ is determined by the projection weight of $\psi_{\pm,k}$ to each of the layers. In this case, the phase factor in the tunneling integral must be derived from the symmetry, like $F = e^{i\phi}$ with $\phi =\arg{(1+e^{iG_1\cdot d}+e^{iG_2\cdot d})}$ for a $K$ band extrema, or $F = 1$ for a $\Gamma$ point band extrema in a hexagonal lattice~\cite{wu2019topological,bistritzer2011moire}.
The second method is more straightforward, but computationally more intensive.
The $c$ coefficients are calculated by taking the inner product of the bilayer Bloch wave functions with a single (top/bottom) layer projected Bloch function (derived from a separate ML DFT calculation and shifted to the corresponding position of the bilayer) at the same band edge momenta:
\begin{equation}
c_t^\pm= \braket{\phi_{k}^t(r) | \psi_{k}^{\pm}(r)}, c_b^\pm= \braket{\phi_{k}^b(r)|\psi_{k}^{\pm}(r)}
\end{equation}
Although the derived coefficients $c$ still have a phase factor uncertainty due to the gauge-freedom of the Bloch phase, that factor is the same for $c_b^+$ and $c_t^+$ and will be canceled when calculating $\Delta_T$.

After deriving the Hamiltonian from the DFT, the next step is to solve the moir\'e electronic structures. The Bloch wave functions across the moir\'e superlattice are
\begin{equation}
\psi_{n\vec{k}}(\vec{r})=e^{i\vec{k}\cdot \vec{r}}\left( \begin{matrix} u_{n\vec{k},t}(\vec{r}) \\ u_{n\vec{k},b}(\vec{r})\end{matrix}\right),  
u_{n\vec{k},t/b}(\vec{r})=\sum_m \gamma_{n\vec{k},t/b}^{m}e^{i\vec{G}_m\cdot \vec{r}}
\end{equation}
where $m$ runs over all reciprocal lattice vectors ($\vec{G}_m$) of the MBZ.
We then derive the matrix form of the Hamiltonian 
\begin{equation}
H(\vec{k}) =
\left( \begin{matrix}
\lambda_{\vec{k}}+\Delta_{t}(\vec{G}_{u}-\vec{G}_{v}) & \Delta_{T}^*(\vec{G}_{u}-\vec{G}_{v}) \\ \Delta_{T}(\vec{G}_{u}-\vec{G}_{v}) &\lambda_{\vec{k}} +\Delta_{b}(\vec{G}_{u}-\vec{G}_{v}) 
\end{matrix}\right) 
\label{eq:matrixH}
\end{equation}
with $\lambda_{\vec{k}} = 
-\hbar^2(k-k_0+G_u)^2\delta_{uv}/2m^*$. To derive the moir\'e electronic structure, we solve the eigenvalue problems of the effective Hamiltonian matrix in Eq.~\ref{eq:matrixH}.
The MBZ reciprocal vectors $G_m$ represent the truncated basis for the continuum expansion about $k_0$.
We find that a $|G_m| < 7 |G_\textrm{sc}|$ is a sufficient number of Fourier components to yield convergent eigenenergies.
The $\Delta$ potentials have been moved to Fourier space by a 2D fast Fourier transform (FFT), but the grid-sampling of the moir\'e potentials are first up-scaled to a $27 \times 27$ real space grid (divisible by 3 to capture the $AB$/$BA$ stacking points and to improve the accuracy) by using a harmonic expansion of the original $9 \times 9$ grid.
The relaxations are also computed in real space on a $27 \times 27$ grid, and the updated potentials (given by $\Delta(\vec{d}_0(r) + \vec{u}(r))$) are properly symmetrized before applying the FFT.
The Chern number of each bands can then be computed as the integral of the Berry curvature through the 1$^{st}$ Brillouin zone \cite{RevModPhys.83.1057}. 

\begin{table*}
\caption{Coefficients of the band extrema properties. Most band extrema locate at high-symmetry $k$-points of K, $\Gamma$, and M. If there are two competing band extrema, we denote $K_1$/$K_2$ ($K_1$) where $K_1$ is the band extrema for untwisted structures, and the following data is for $K_1$ edge. $K_1-K_2-K_3$ means the band extrema locate on the high symmetry line between the referred high symmetry points. $m^*, V$, and $w$ denote the effective mass, the 1$^{st}$-order Fourier coefficients of electrostatic potential and tunneling matrix elements. $\theta_c$ is the critical twist angle for the electronic localization when the width of the uppermost band is 10 meV.}
\begin{tabular}{|c|c|c|c|c|c|c|} 
\hline
Materials & Band & Extrema & \ \ \ $m^*$ ($m_e$) \ \ \ & \ \ \ $V$ (meV)\ \ \ &\ \ \ $w$ (meV)\ \ \ & \ \ \ $\theta_c$ ($^0$)\ \ \ \\ 
\hline 
MoS$_2$ & conduction & K & 0.54 & 5.64 & 4.8 & 1.95 \\ 
 & valence & $\Gamma$ & 2.6 & 12.01 & 95.8 & 4.11 \\ 
\hline 
MoSe$_2$ & conduction & K & 0.53 & 6.48 & 7.12 & 2.07 \\ 
 & valence & $\Gamma$ & 5.74 & 20.29 & 107.96 & $>$6 \\ 
\hline 
MoTe$_2$ & conduction & K & 0.57 & 8.01 & 4.57 & 2.30 \\ 
 & valence & K & 0.68 & 9.34 & 10.3 & 2.81 \\ 
\hline 
WS$_2$ & conduction & K & 0.34 & 2.02 & 1.52 & 1.48 \\ 
 & valence & $\Gamma$ & 3.21 & 2.67 & 64.22 & 4.85 \\ 
\hline 
WSe$_2$ & conduction & K & 0.35 & 9.18 & 8.75 & 1.70 \\ 
 & valence & K/$\Gamma$ (K) & 0.46 & 8.55 & 11.1 & 2.14 \\ 
\hline 
WTe$_2$ & conduction & K & 0.32 & 6.85 & 7.28 & 1.74 \\ 
 & valence & K & 0.45 & 10.27 & 11.4 & 2.34 \\ 
\hline 
hBN & conduction & K/M (K) & 0.69 & 99.21 & 134.85 & 3.48 \\ 
 & valence & K & 0.52 & 45.42 & 58.53 & 2.55 \\ 
\hline 
SiH & conduction & $\Gamma$ & 0.23 & 1.35 & 3.5 & 1.34 \\ 
 & valence & $\Gamma$ (degenerate) & -- & -- & -- & -- \\ 
\hline 
InSe & conduction & $\Gamma$ & 0.32 & 15.3 & 138.37 & 1.70 \\ 
 & valence & $\Gamma$ (degenerate)& -- & -- & -- & -- \\ 
\hline 
BP & conduction & $\Gamma$ & 1.18 & 26.77 & 174.95 & 3.44 \\ 
 & valence & $\Gamma$ & 3.09 & 19.93 & 109.75 & 1.57 \\ 
\hline 
Ars. & conduction & $\Gamma$ & 2.01 & 37.55 & 249.52 & $>$6 \\ 
 & valence & $\Gamma$ & 1.95 & 29.62 & 156.15 & $>$6 \\ 
\hline 
GaS & conduction & $\Gamma$ & 0.29 & 11.78 & 67.93 & 1.37 \\ 
 & valence & K-$\Gamma$-M & -- & -- & -- & -- \\ 
\hline 
GaSe & conduction & $\Gamma$ & 0.21 & 15.59 & 91.95 & 1.25 \\ 
 & valence & K-$\Gamma$-M & -- & -- & -- & -- \\ 
\hline 
\end{tabular}
\label{tab:elect}
\end{table*}
\section{Relaxation and Electronic Coefficients of Materials} \label{sec:coef}
Applying our algorithm to 18 different twisted bilayer materials, we derive the coefficients for both atomic relaxation and electronic structures. The coefficients for mechanical properties related to atomic relaxation are shown in Fig \ref{tab:relax}. Strain modulus K and G are fitted as $E=\int \frac{1}{2}K(u_{xx}+u_{yy})^2+\frac{1}{2}G[(u_{xx}-u_{yy})^2+4u_{xy}^2]$, representing the isotropic and anisotropic strain modulus, respectively. The GSFE energy is expanded as the Fourier series: GSFE(r)$=\sum_{u,v}F_{uv}e^{i(ub_1+vb_2)\cdot r}$. For materials with hexagonal lattices, the 1$^{st}$ order components of $F_{ij}$ are equal and we adopted two times of them as $c_1$ to represent the GSFE intensity.\cite{carr2018relaxation} For materials with other lattices, we use $\frac{2}{9}$(max GSFE -min GSFE) for comparison, as this is consistent with the previous definition for hexagonal lattices. The relative intensity of the GSFE and strain energy determines the characteristic twist angle when the in-plane relaxation effects play important role. Large GSFE leads to strong in-plane relaxation, or namely, large upper-limit of the twist angle where the relaxation plays dominant role on the atomic structure. According the analysis in previous work \cite{PhysRevB.96.075311}, the width of the domain wall $w_d$ and critical twist angle $\theta_c$ are estimated as follow:   
\begin{equation}
w_d=\frac{a}{4}\sqrt{\frac{K}{c_1}},   	\theta_c = \frac{4}{\sqrt{3}}\sqrt{\frac{c_1}{K}}
\label{equ:ang_rel}
\end{equation}
where a is the unit cell parameter (of hexagonal lattices). We set the criterion of $\theta_c$ as the condition when the domain walls overlap at $AB$/$BA$ stacking.
\begin{figure*}[hbtp]
\centering
\includegraphics[scale=0.36]{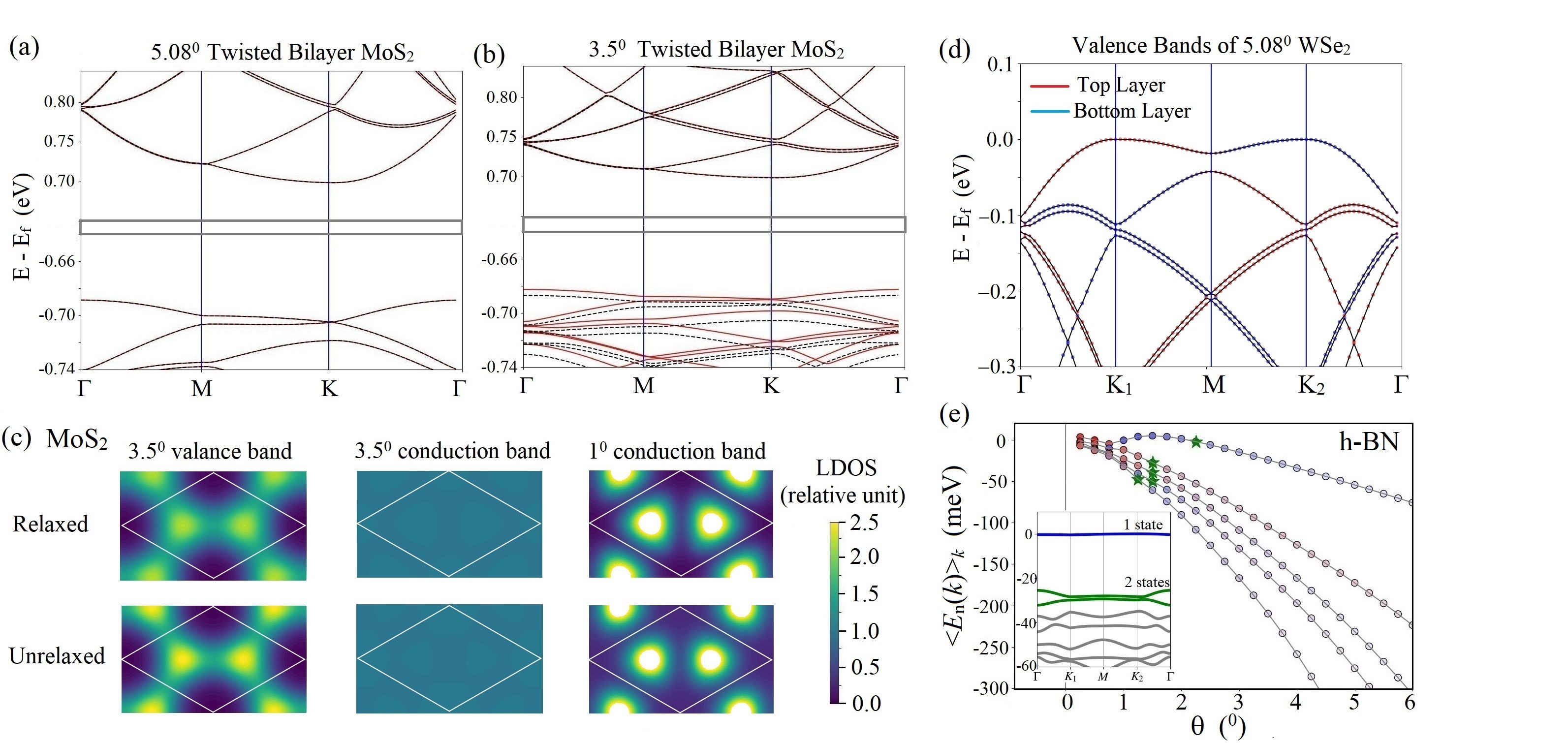}
\caption{Moire electronic properties of TB MoS$_2$. 
Band structures of (a) $5.08^{\circ}$ and (b) $5.08^{\circ}$ MoS$_2$. The VBM and CBM derive from $\Gamma$ and K-edges, respectively. The red lines represent the results with full relaxation, and the black dash lines represent the results without in-plane relaxation.
(c) Realspace distribution of the MoS$_2$ conduction and valance bands represented by the local density of states (LDOS) with and without the in-plane relaxation. The LDOS are normalized to have the average value of 1 in the MSL and plotted with the same color scale from 0 to 2. (d) Valence bands of twisted bilayer WSe$_2$. The red and blue dots are the projection weight to the top and bottom layers. (e) Angle-dependence of moir\'e bands for twisted bilayer h-BN. The format of this panel is the same as Fig. 3b in the main text. The small panel is the band structure at $\theta = 1^0$.}
\label{fig:compare}
\end{figure*}

The energy coefficients of these materials are also plotted in Fig \ref{fig:matlist}a to visualize their comparison. The materials can be generally divided into three class: the materials with high strain modulus and low GSFE coefficients (graphene, h-BN, SiH) are stiffer during relaxation, and their characteristic angles $\theta_c$ are about $1\degree$; the materials with the intermediate proportion between the two coefficients (TMDCs, BP, etc.) have $\theta_c$ in the range of $1-3\degree$; while the materials with high GSFE coefficients and low strain modulus (InSe, Asenene, Bi) have low stiffness and displays evident relaxation around or above $3\degree$. This estimation is also well confirmed by the computed results of relaxation.

The electronic properties of twisted bilayers are divided into Moire molecule and Moire crystals according to the degree of real space localization\cite{carr2018relaxation}, which is determined by the relative importance of the kinetic and potential energy in the MBZ. The comparison of the effective mass and band edge energy fluctuation in configuration space are visualized in Fig \ref{fig:matlist}b. As the kinetic energy scales with $\theta$ as $T\propto \theta^{2}(m^*)^{-1}$, electronic localization appears in the low $\theta$ region where $\theta < \frac{a}{\pi \hbar}\sqrt{m^*V}$. The critical twist angles for each materials when the uppermost band width is about 10 meV are listed in the table. The h-BN, \uppercase\expandafter{\romannumeral5} group elementary substances and $\Gamma$-edge TMDs typically have a high ratio of the potential to kinetic energy coefficients, indicating the evident localization appears for $\theta$ at about $3-4 \degree$; while the common TMDs with K-edge and some other materials (like InSe, SiH, ect.) display significant localization only for small twist angle of about $1-2\degree$. Otherwise, the product of the two coefficients determines the frequency $\omega_{\theta}$ in the harmonic oscillator model\cite{carr2020duality}, which determines the slope of the average energy of the top bands to the twist angle. 

The complete data is listed in Tab~\ref{tab:elect} including the band extrema and relative intensity of electrostatic potential, tunneling, and kinetic energy (represented by effective mass). Specifically, the band extrema at $\Gamma$ point typically host stronger tunneling and energy fluctuations, as well as a larger scale of bandgap variation during the inter-layer displacement and twisting.

Our present calculations have not covered the degenerate band edge or band extrema out of the high symmetry $k$-points. However, there is no substantial obstacle for our methods to be generalized to different $k$-points.

\section{Consistency with Previous work} \label{sec:matlist}
\begin{figure*}[hbtp]
\centering
\includegraphics[scale=0.27]{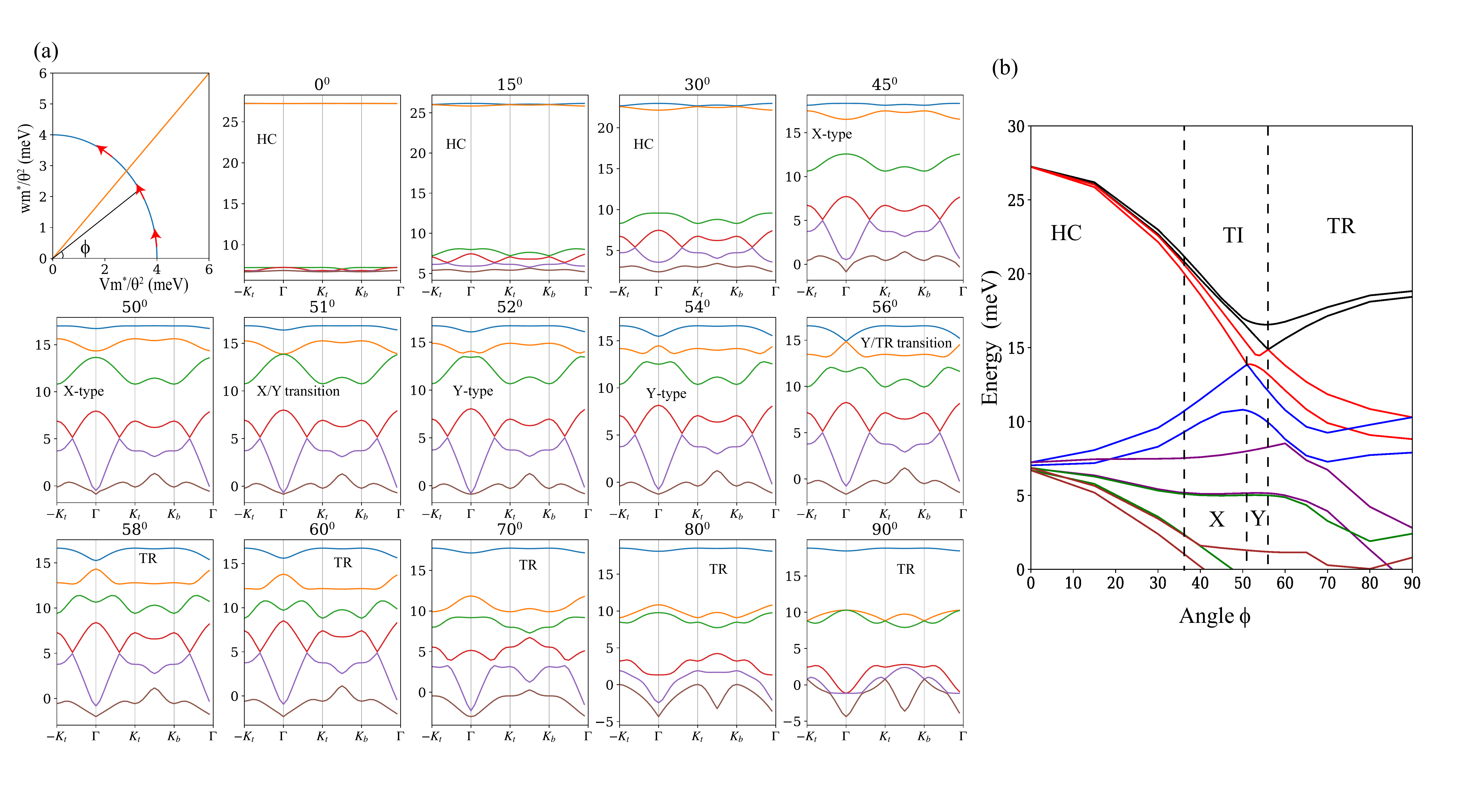}
\caption{Band reordering during the transition from TL to HL QHO conditions. (a) Calculated moire band structures for the points on the 1/4 circle in the phase diagram with a radius of 4 meV and twist angle $\phi$ as shown in the left-top panel. The angle $\phi$ gradually changes from 0$^0$ to 90$^0$ and the band structures change from HL, A-type, B-type, to TL condition through the closing and reopening of the band gaps. (b) Evolution of the moire band maximum and minimum as a function of $\phi$. The 1$^{st}$, 2$^{nd}$, 3$^{rd}$, and 4$^{th}$ bands are denoted by black, red, blue, and purple lines, respectively. For each color, the upper line is the band maximum and the lower line is the band minimum. }
\label{fig:reorder}
\end{figure*}
To validate our methods, we also do calculations that could compare with the previous full DFT calculations, as shown in Fig \ref{fig:compare}. The bands of 3.5$^0$ and 5.08$^0$ twisted bilayer MoS$_2$ in Fig \ref{fig:compare}a,b are consistent with the full DFT results in ref \cite{naik2018ultraflatbands}. The real-space localization of the uppermost band states at $AB$/$BA$ stacking in Fig \ref{fig:compare}c is also the same with the full DFT results of the relaxed system. Valence bands of 5.08$^0$ twisted bilayer WSe$_2$ shown in Fig \ref{fig:compare}d are accurately consistent with ref \cite{wang2020correlated} regarding both the band energy and projection weight to the top and bottom layer. This panel is also a good example to show how the moir\'e bands originate from the ML bands as illustrated in Fig 2b in the main text. The angle-dependent bands of twisted bilayer h-BN are shown in Fig \ref{fig:compare}e. We derive the consistent results of the pairing behavior (single uppermost band and pairing 2$^{nd}$ and 3$^{rd}$ bands) and angle dependence with the moir\'e bands in ref \cite{xian2019multi}.

\section{Bands Reordering During the Topological Phase Transition} \label{sec:reorder}
Here we show the band reordering process during the topological transition as illustrated in Fig 1d in the main text. To present the change from $V$ dominant to $w$ dominant moir\'e potential, we select a 1/4 circle route in the phase diagram as shown in the first panel of Fig \ref{fig:reorder}a. On this route, the kinetic energy term keeps constant, and only the relative intensity of $V$ and $w$ changes, as represented by the variation of $\phi$. The HL QHO bands appear at the small $\phi$ region. When tuning up $\phi$, the two top bands separate at K-point at about $\phi =30^0$ to form the A-type topological phase. And then, the third band lifts and contact with the second band at $\Gamma$-point at $\phi = 51^0$, exchanging the Chern number by 2 and experiencing phase transition to B-type topological phase. The second band then contacts with the first band at $\Gamma$-point at $\phi = 56^0$ and exchange the Chern number by 1, making the Chern number of the uppermost band zero. The system finally enters the TL phase where the topologically trivial uppermost bands separate from the lower bands.

This process is cohesively shown in Fig \ref{fig:reorder}b, where the maximum and minimum energy of each band are plotted against the parameter $\phi$. Ideal band degeneracy sequences of (1,2,3,...) and (2,4,6,...) display on the right and left side, respectively. Their reordering leads to the connection and separation in the intermediate region, where the TI phases appear.

\bibliography{refs.bib}